%% file: Main.tex
\documentclass[conference]{IEEEtran}
\IEEEoverridecommandlockouts

\usepackage{cite}
\usepackage{booktabs} 
\usepackage[listings]{tcolorbox}
\usepackage{xcolor}
\usepackage{multirow}
\usepackage{hyperref}
\usepackage{url}
\usepackage{comment}
\usepackage{todonotes}
\usepackage{color}
\usepackage{balance}
\usepackage{adjustbox}
\usepackage{tabularx}
\usepackage{amsmath,amssymb,amsfonts}
\usepackage{algorithmic}
\usepackage{graphicx}
\usepackage{textcomp}
\usepackage{xcolor}
\def\BibTeX{{\rm B\kern-.05em{\sc i\kern-.025em b}\kern-.08em
    T\kern-.1667em\lower.7ex\hbox{E}\kern-.125emX}}
\begin{document}

\title{Analysis of Software Engineering Practices in General Software and Machine Learning Startups
}

\author{\IEEEauthorblockN{Bishal Lakha}
\IEEEauthorblockA{\textit{Computer Science Department} \\
\textit{Boise State University}\\
Boise, ID, USA \\
bishallakha@u.boisestate.edu}
\and
\IEEEauthorblockN{Kalyan Bhetwal}
\IEEEauthorblockA{\textit{Computer Science Department} \\
\textit{Boise State University}\\
Boise, ID, USA \\
kalyanbhetwal@u.boisestate.edu}
\and
\IEEEauthorblockN{Nasir U. Eisty}
\IEEEauthorblockA{\textit{Computer Science Department} \\
\textit{Boise State University}\\
Boise, ID, USA \\
nasireisty@boisestate.edu}

}

\maketitle

\begin{abstract}
\input{abstract.tex}
\end{abstract}

\begin{IEEEkeywords}
Software Engineering, Machine Learning Startups, Software Startups, Systematic Literature Review
\end{IEEEkeywords}

\section{Introduction}
\label{sec:Introduction}
\input{sec_introduction}

\section{Background}
\label{sec:Background}
\input{sec_background}

\section{Research Methodology}
\label{sec:Methodology}
\input{sec_methodology}

\section{Results}
\label{sec:Results}
\input{sec_results}

\section{Threats to Validity}
\label{sec:Threats}
\input{sec_threats}

\section{Discussion and Conclusion}
\label{sec:Discussion}
\input{sec_discussion}

\bibliographystyle{abbrv}
\bibliography{sigproc}  

\end{document}

%% file: abstract.tex
\textit{Context:} On top of the inherent challenges startup software companies face applying proper software engineering practices, the non-deterministic nature of machine learning techniques makes it even more difficult for machine learning (ML) startups.
\textit{Objective:} Therefore, the objective of our study is to understand the whole picture of software engineering practices followed by ML startups and identify additional needs.  
\textit{Method:} To achieve our goal, we conducted a systematic literature review study on 37 papers published in the last 21 years. 
We selected papers on both general software startups and ML startups. We collected data to understand software engineering (SE) practices in five phases of the software development life-cycle: requirement engineering, design, development, quality assurance, and deployment. 
\textit{Results:} We find some interesting differences in software engineering practices in ML startups and general software startups. 
The data management and model learning phases are the most prominent among them. 
\textit{Conclusion:} While ML startups face many similar challenges to general software startups, the additional difficulties of using stochastic ML models require different strategies in using software engineering practices to produce high-quality products. 

%% file: sec_introduction.tex
Machine learning is becoming ubiquitous in many software applications. Startups and small companies are eagerly adopting this technology. They are the flag bearers for implementing innovative and state-of-the-art ML solutions for different domains \cite{spender2017startups}. The challenging nature of ML application and the limitations and peculiarities of startups have resulted in using slightly different software engineering practices. However, such systems' development, deployment, and maintenance still suffer from the lack of best practices. 

Due to the non-deterministic nature of machine learning, all software engineering aspects for ML systems become complicated \cite{giray2021software}. They also lack proper and mature tools to test ML systems \cite{giray2021software}. So integrating ML in software applications has forced organizations to evolve their development process \cite{amershi2019software}. In a research conducted by Amershi et al. \cite{amershi2019software}, they studied methods followed by different software engineering teams developing artificial intelligence (AI) products at Microsoft. Their study suggested that AI components are inherently complex than other software components, and while integrating them, non-monotonic error behavior can arise. They proposed end-to-end pipeline support, data collection, cleaning, and management tools, focusing on programming and model bugs, and others to mitigate such problems. A few companies have recently adopted new approaches like MLOps, but there still are many challenges \cite{9356947}. These challenges become more prominent in startups. 

More than 60\% of startups do not survive their first five years, and 75\% of the startups which venture capitalists back meet failure\cite{nobel2011companies} while more than 90\% of startups go bankrupt \cite{eloranta2014patterns}. One of the crucial reasons for such situations is improper software engineering practices resulting in faulty products \cite{crowne2002software}. Moreover, many startups take more than a year to develop a minimum viable product, and due to delay in the product delivery, additional financial burden occurs resulting in startup failure \cite{kalyanasundaram2018startups}. Similarly, the inability of organizations to actively engage developers also results in the failure \cite{akter2020failure}. Again, these manifestations are consequences of bad software engineering practices.

Malpractices in software engineering do not always lead to startup failure. Other factors include different technical debt like code debt, architecture debt, or testing debt. These debts could accumulate rapidly, impacting the performance and growth of a startup \cite{8449238}. AI/ML startups tend to fall more than their traditional counterparts, and the cause mentioned above also plays a role in their failure. That is why learning about what software engineering practices they follow and which best practices are crucial.

Various studies have been conducted about software engineering practices in the tech industry. Also, there are many studies conducted on software engineering practices in startups. However, studies on ML startups and their software engineering practices are scarce. This scarcity motivated us to direct our study in that area. Moreover,  it occurred to us that since the ML field is evolving rapidly, the changes in software engineering practices in those companies, especially in startups, should be studied along with their difference from general software startups.

Therefore the main objectives of our study are:
\begin{itemize}
    \item Identify the key software engineering practices followed by software startups in general
    \item Identify specific software engineering practices followed by ML startups
    % \item The practices or traits seen in successful machine learning startups
\end{itemize}

%% file: sec_background.tex
The emergence of different electronic devices like smartphones, tablets, laptops, smartwatches, etc., has contributed to the software industry's manifold growth, resulting in a multitude of software startups \cite{buganza2015unveiling}. The success of deep learning, a sub-field of ML, in a wide range of applications resulted in the adoption of ML  in multiple companies and pushed the growth of startups rapidly \cite{schulte2021scaling} leading to billions of dollars of contribution to the economy\cite{chui2017artificial}.

\textbf{General Software Startups.}
Software startups small or medium-sized enterprises distinct from traditional mature companies which focus on developing innovative products in a limited time frame and resources \cite{unterkalmsteiner2016software}. Due to their inherent characters, they face multiple challenges like little organization management experience, lack of financial and human resources, influences from various sources like investors, customers, partners, and continuous need of developing dynamic and disruptive technologies \cite{854066}.

\textbf{Machine Learning Startups.}
Machine learning startups provide different ML  services like speech recognition, video analysis, and others or use ML  as a part of their product. Many startups of different domains like health\cite{garbuio2019artificial} \cite{vijai2021rise}, finance \cite{8614276}, fashion \cite{luce2018artificial}, etc. are also adopting AI and ML  in their product.  While developing such products, startups face multiple uncertainties due to the gap between research and development \cite{8486814}. As software startups are more prone to failure \cite{cantamessa2018startups}, these kinds of uncertainties make ML startups more vulnerable. However, collaboration, cooperation, and openness, which come from good software engineering practices, can help such startups succeed \cite{blake2020success}.

\subsection{Research questions}
To fulfill our objectives, we formulated two research questions and conducted a systematic literature review of multiple published papers in different venues.

\textbf{RQ1: Which software engineering practices are followed by general software startups?}

With RQ1, we intend to find which SE practices do general software startups follow and how they are similar or different from traditional SE practices. These practices will illustrate the characteristics of software startups along with reasons for their choices of particular software design patterns and development practices.  

\textbf{RQ2: Which additional software engineering practices do machine learning startups follow?}
		
With RQ2, we intend to differentiate general startups from ML startups using SE practices. We also plan to learn different steps involved in ML product development, what general practices still hold their utility, and what modifications, improvements, and innovations are necessary to address those steps.

%% file: sec_methodology.tex
We collected published papers and manually went through all the selected papers to collect data. To maintain the quality of our study, the first and second authors individually went through each of the papers. After collecting the data, we analyzed and synthesized the data to answer our research questions. In this section, we discuss our research methodology in detail. 

\subsection{Inclusion and Exclusion Criteria}
We have included papers published only in English and considered papers published in journals, conferences, or symposium proceedings for general software startups. Since published papers on 
 ML  startups were rare, we included a few blogs. We avoided other gray literature like videos, podcasts, news articles, interviews, etc. Also, we only considered papers published until November 2021 for this systematic literature review. 

\subsection{Database Search}
We formulated our search strings to download papers to fulfill our research objective. We then searched papers in IEEE Xplore, ACM Digital Library, and Google Scholar using our formulated search strings. At first, we used two search strings on those online databases. The query 1 is ``Software Engineering and Startups". The query 2 is ``ML and Software Engineering and Startups". There was a high number of papers returned but only few were relevant. We further searched on Google Scholar to enrich our database of papers.

\begin{comment}
\begin{table}[ht]
 \centering
 \caption{Paper Counts for Search in each database}
 \begin{tabular}{l|c|c}
 \hline
 \hline
 Database & Query 1 &  Query 2\\
 \hline
 IEEE Xplore & 389 & 13\\
 \hline
 ACM & 12274 & 10040\\
 \hline
Google Scholar & 9640 & 2380\\
 \hline
\end{tabular}
\end{table}
\end{comment}

We further formulated sub-queries based on each stage of the software development life-cycle (SDLC) for general software startups. In Query 3, we searched ``Software Engineering and Startups and Requirement Analysis". In Query 4, we searched for ``Software Engineering and Startups and Design". In Query 5, we searched for ``Software Engineering and Startups and Implementation and sdlc". In Query 6, we searched for ``Software Engineering and Startups and testing and sdlc". In Query 7, we searched for ``Startups and Deployment and sdlc". Then we downloaded the meta-data for further processing.

\begin{comment}
 \begin{table}[ht]
 \centering
  \caption{Paper Counts for Search in each database for each phase of SDLC in software Startups}
 \begin{tabular}{c|c|c|c|c|c}
 \hline
 \hline
 Database & Q3 &  Q4  &  Q5  &  Q6 & Q7\\
 \hline
 IEEE Xplore & 14 & 16 & 10 & 10 & 5 \\
 \hline
 ACM  & 10,374 & 11,167 & 17 & 22 & 11\\
 \hline
Google Scholar & 200 & 8430 & 249 & 233 & 160\\
 \hline
\end{tabular}
\end{table}
\end{comment}

%\begin{figure}
%  \centering
 % \includegraphics[width=1.0\linewidth]{photos/SE_Methodology.png}
%  \caption{Research Methodology Flow Diagram}
%  \label{research_methodology}
%\end{figure}

Our goal was to collect papers for ML  startups based on software development life cycles; we did a similar query search in IEEE Xplore and ACM Digital Library but did not find relevant papers. We then use alternate terminologies for machine learning like ``Deep learning" and ``Artificial Intelligence" to enrich our database for ML  startups and got few additional useful papers.

\subsection{Meta Data Collection}
We collected metadata -  title, author names, abstract, published year, URL, citations, and other - of the papers from the database search in a CSV file. We did this to identify deduplication and validation of papers. This collection also allows the filtering of the papers based on abstracts manually. IEEE explorer has inbuilt features that enable us to export metadata of all search results as a CSV file. However, those features were not available in ACM Digital Library and Google Scholar. So, we built a web scraper using the python library BeautifulSoup and extracted the metadata from ACM Digital Library search results. 

\subsection{De-duplication and Validation}
We collected the metadata of 389 papers for general software startups from IEEE explorer, 12,274 papers from ACM Digital Library, and 9,640 papers from google scholar. Similarly, we collected 13 papers for ML startups from IEEE explorer, 10,040 papers from ACM Digital Library, and 2,380 papers from google scholar. We also collected 39 papers curated in a site ml-ops.org. We then dropped all the duplicate papers from the list using the metadata. We used a python package called Pandas for this work.  

\begin{figure}
  \centering
  \includegraphics[width=1.0\linewidth]{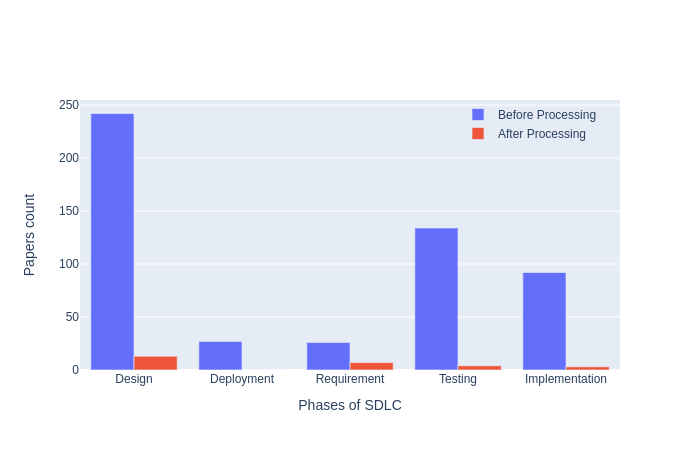}
  \caption{Number of papers per Software Development Life-Cycle phases before and after filtration}
  \label{before_and_after_filter}
\end{figure}

We also noticed that the abstract of many papers collected from the search doesn't have the keywords used to search them. In order to remove such papers, we used RegEx. It uses a sequence of characters in order to find a specific pattern in the text. We used it to search if all the required keywords were present or not in the given abstract. We dropped the paper from our list if we didn't find one or more keywords in the abstract. Figure~\ref{before_and_after_filter} shows the number of papers belonging to different software development life-cycle for general software startups before and after filtration based on the presence of keywords. 

\subsection{Snowballing}
Our database search for general software startups resulted in enough papers for our research. However, we got very few papers for ML  startups. We, thus, used snowballing techniques to increase our database for ML  startups. We went through the citations of the papers we found and checked if any of those were relevant to us based on our inclusion-exclusion criteria. We previously had sixteen relevant papers, and after snowballing, we got twenty-one papers. 

\subsection{Manual Selection and Finalization}
After de-duplication and validation, we had 92 total papers, where 72 papers belonged to general software startups, and 20 papers belonged to ML startups. The automatic filtration did a good job in removing irrelevant papers, but several papers contained required keywords but were not useful for our research. So we went through the abstract of all the remaining papers and manually selected papers. We finally got 37 papers, 27 for general software startups and 10 for ML startups distributed among different phases of SDLC as shown in Table ~\ref{tab:paper_count}{}.

 \begin{table}[ht]
 
 \centering
  \caption{Final paper count based on SDLC for General and ML Startups}
%   \resizebox{\columnwidth}{4}{8}{%

 \begin{tabular}{c|c | c}
 \hline
 \hline
 Phase & General Startup &  ML Startup   \\
 \hline
Requirement Engineering & 5 & 3\\
 \hline
Design & 7 & $\times$ \\
 \hline
Data Management & $\times$ & 4\\
 \hline
Development & 6 & $\times$\\
 \hline
 Model Learning & $\times$ & 4 \\
 \hline
 Testing & 7 & 5 \\
 \hline
 
 Deployment & 2 & 5 \\
 \hline

\end{tabular}
\label{tab:paper_count}

% }
\end{table}

%  \begin{table}[ht]
%  \centering
%   \caption{Final papers count based on SDLC for ML Startups}
%  \begin{tabular}{c|c}
%  \hline
%  \hline
%  Phase & Count   \\
%  \hline
% Requirement Engineering & 3\\
%  \hline
% Data Management & 4\\
%  \hline
% Model Learning & 4 \\
%  \hline
%  Testing & 5\\
%  \hline
%  Deployment & 5\\
%  \hline
% \end{tabular}
% \end{table}

%% file: sec_results.tex
We present our findings per research questions in this section. 

\subsection{RQ1: Which software engineering practices are followed by general software startups?}

We summarized the various practices followed by general software startups based on SDLC in Table \ref{tab:general_startup_summary}{}  and discussed them in detail in the following subsections. We also discuss potential challenges and solutions found in the study.

\subsubsection{Requirement Engineering}
 It is the first part of the software development process. A well-placed requirement can help solve problems better and develop efficient software with the best quality. In this phase of SDLC, vague high-level user requirements are translated into complete, precise, and formal specifications for the further development of software \cite{Chakraborty}. It involves interaction between both the producers and end users of the software. There are various approaches for requirement engineering. Startups also employ different techniques based on their needs.

Most of the time, startups struggle not knowing what they should develop because software requirements are not clearly mentioned \cite{9026036}. Melegati et al. \cite{9026036} conducted a total of nine interviews with founders or managers from various Brazilian startups using a grounded theory approach about requirement engineering practices in startups. They found that founders or managers play vital roles in selecting practices in startups. They also observed different levels of maturity in requirement engineering in software development. They found that some startups did not have any formal process; some startups had clear steps for requirements engineering. In cases where there were matured requirement practices, it mainly occurred due to market pressure or requirements. The requirement engineering differs depending on the nature of the startup, such as client-based or user-target-based startups. If it is a client-based startup, the client sets all the requirements. But if it is a user-target-based startup, the owners set the requirements. In both cases, a validation step takes place to check for the viability of the requirement. There can be no or little software development in this phase. After development is completed, a final validation step is taken to see if the requirements were correctly implemented. 

Rafiq et al. \cite{8051340} studied three startups around the globe about the requirement elicitation process. They found that requirement engineering in startups is very primitive and informal, and it continues to develop alongside product development. The owners already have something in mind before the requirements engineering process begins. Then it matures as time evolves. The most common requirement elicitation techniques are conducting interviews, prototyping, and brainstorming. In addition, some unconventional methods such as competitor analysis, collaborative team discussions, and model users are also used. 

Alves et al. \cite{9218221} conducted a literature review on requirement engineering in startups. They found that startups use flexible and informal requirement engineering. In addition, they observed that startups are more concerned with evolution by acquiring more customer base using pragmatic requirement practices. They also conducted a case study on ten startups based on the Digital Port ecosystem. They concluded that startups use straightforward requirement engineering even they mature by growing the customer base.

Gralha et al. \cite{8453156} studied the evolution of requirements engineering in 16 software startups using a grounded theory approach. They found six key factors that evolve relevant to requirement engineering: requirements artefacts, knowledge management, requirements-related roles, planning, technical debt, and product quality. Furthermore, they found that advances in one dimension often facilitate advances in other dimensions. But the interesting conclusion is that evolution in these dimensions is not fundamental to the success of the startups. But they have a positive impact on the product, employee, and company as a whole.

\subsubsection{Design}
Design is the 2nd phase of software development and comes right after requirement engineering. In this phase, software architects and developers design a high-level system architecture based on requirements. Some software startups follow design principles and methodology, while some might not. 

Startups present a founder-centric approach, and depending on founders' background; they might encourage some architectural design before the development phase \cite{7360225}. Since founders are the ones who have taken the risk, they have the upper hand in deciding the design and other considerations. If startups fail to generate revenue, they face enormous consequences.

Crowne et al. \cite{1038454} found that startups don’t have experienced developers, so they neglect non-coding issues like architecture and design. Startups also have financial constraints. So, it is challenging for them to hire experienced and quality human resources that directly affect the architecture and design of the software.

Duc et al. \cite{duc2016minimum} performed an empirical study on five early-stage startups based on interviews, observation, and documentation. They found that startups were unaware of Minimum Viable Product (MVP). MVPs facilitate cost-effective product design and bridge the communication gap. 

Deias et al. \cite{deias2002introducing} found that there is a lack of well-written architectural and design specifications in startups. This lack is partly due to time and resource constraints. In addition, startups have pressure to deliver products as soon as possible, leading to design practices being compromised. 

Souza et al. \cite{10.1145/3350768.3350786} observed that all of the startups they studied construct a simple design of software in a quick session with their customers since they closely work with them.

Paternoster et al. \cite{paternoster2014software} in their systematic study concluded that the use of well-known frameworks supports rapid product change. Also, Jansen et al. \cite{4670713} in their study on two startups, found that opportunistic and pragmatic reuse of third party software helped in the rapid development of software, hence reducing time to market. In addition, code reuse reinforces the architectural structure of the product and increases the product's ability to scale.

\subsubsection{Development}
In this phase of software development, requirements and design are implemented into system components. Developers write code and design databases along with IT infrastructure to support them. Startups follow various development practices to implement their product.  

NicolòPaternoster et al. \cite{paternoster2014software} conducted a systematic mapping study on software development practices in startups. They found that startups don’t follow any standard software development practices. This tendency is justifiable because startups are primarily concerned with delivering their products in the market as early as possible to start revenue generation. In addition, they want to minimize the cost of development. Therefore, both time and capital need to be invested in establishing a formal process. But both of these are significant constraints for startups. Therefore, startups mostly go after unpredictable, responsive, and low precision software engineering \cite{854066} \cite{kakati2003success}.  

Heitlager et al. \cite{4383096} found that startups generally share a common pattern: few individuals starting with scarce resources. Coleman et al. \cite{coleman2008investigation} observed the same. Startups have minimal resources and only use their limited resources to support product development. 

Dande et al. \cite{dande2014software} studied working practices in startups in Finland and Sweden. They found 63 common practices used by software startups.

Souza et al. \cite{10.1145/3350768.3350786} studied agile development in software startups by conducting 14 interviews with the CTO and CEO of startups. They found that tools and processes backed most development activities to facilitate the software development process. For example, using a version control system enables the continuous integration and deployment of software.

% \usepackage{booktabs}
% \usepackage{graphicx}
\begin{table*}[]
\centering
% \resizebox{\textwidth}{!}{%
  \begin{adjustbox}{width=\textwidth}
{\large
\begin{tabular}{lllll}
\toprule
Knowledge Area          & Primary Study ID                                                                                                                                                                                                         & Summary                                                                                                                                                                                                                                                                                                                                                                         \\ \midrule
Requirement Engineering & \begin{tabular}[c]{@{}l@{}}Chakraborty et  al., 2012; \\ Melegati et  al., 2016;\\ Rafiq et al., 2017;\\Alves  et al., 2020;\\ Gralha et al.,2018;\end{tabular}                                                            & \begin{tabular}[c]{@{}l@{}}Startups generally didn’t have any formal process;\\ Some startups might have clear steps for requirements engineering;\\ In cases where there were matured requirement practices,
it mostly occurred due to market pressure or requirement;\\ Founders play vital roles in what practice to choose;\end{tabular}                            \\ \midrule
Design                  & \begin{tabular}[c]{@{}l@{}}Giardino et al., 2016;\\ Crowne et al., 2002;\\ Duc et al., 2016;\\ Deias et al., 2002;\\ Souza et al., 2019;\\ Paternoster et al.,2014;\\ Jansen et al., 2008;\end{tabular}                  & \begin{tabular}[c]{@{}l@{}}Startups present a founder centric approach\\ Founders are the ones who have taken huge risks, have the upper hand in deciding design\\ Startups don’t have experienced developers so they neglect non-coding issues like architecture and design\end{tabular}                                                                                       \\ \midrule
Development             & \begin{tabular}[c]{@{}l@{}}Paternoster et al.,2014;\\ Sutton et al., 2000;\\ Kakati et al., 2003;\\ Heitlager et al.,2007;\\ Coleman et al., 2008\\ Dande et al., 2014;\\ Souza et al.,2019;\end{tabular}                & \begin{tabular}[c]{@{}l@{}}Most startups don’t follow any standard software development practices\\ Founders want to minimize the cost of development. \\To establish a process, both time and capital need to be invested, which are major constraints for startups. \\ Startups mostly go after unpredictable, responsive, and low precision software engineering\end{tabular} \\ \midrule
Testing                 & \begin{tabular}[c]{@{}l@{}}Shikta et al., 2021;\\ Pompermaier et al., 2017;\\ Giardino et al., 2016;\\ Unterkalmsteiner et al., 2016;\\ Mater et al., 2000;\\ Shikta et al.,2021;\\ Thongsukh et al., 2017;\end{tabular} & \begin{tabular}[c]{@{}l@{}}Startups generally ignore any quality assurance\\ The testing was dependent on the end-user in the first versions\end{tabular}                                                                                                                                                                                                                   \\ \midrule
Deployment                 & \begin{tabular}[c]{@{}l@{}}Silva  et al., 2005;\\ Taipale  et al., 2010;\end{tabular}                                                                                                                                    & \begin{tabular}[c]{@{}l@{}}Some use Manual Deployment\\ Some use CI/CD pipelines\end{tabular}                                                                                                                                                                                                                                                                                   \\ \bottomrule
\end{tabular}%
}
  \end{adjustbox}

% }
\caption{Summary of SE practices in general software startups}
\label{tab:general_startup_summary}
\end{table*}

\begin{table*}[ht]
\centering
% \resizebox{\textwidth}{!}{%

  \begin{adjustbox}{width=\textwidth}

\begin{tabular}{lllll}
\toprule
Knowledge Area          & Primary Study ID                                                                                                                                                                                                         & Summary                                                                                                                                                                                                                                                                                                                                                                         \\ \midrule
Requirement Engineering & \begin{tabular}[c]{@{}l@{}}E.d. S. Nascimento et al. , 2019; \\ A. Banks et al., 2019;\end{tabular}                                                            & \begin{tabular}[c]{@{}l@{}}ML  applications have additional requirements that revolve around training data\\ Creating requirements based on business needs due to the inability of customers to understand \\ data and metrics requirements appropriately is challenging
\end{tabular}                            \\ \midrule
Data Management                 & \begin{tabular}[c]{@{}l@{}}A.Hopkins et al., 2021;\\ A. Arpieg et al., 2018;\end{tabular}                  & \begin{tabular}[c]{@{}l@{}}Different kinds of data bugs appear while preparing training data, and developers use various\\ verification tools.\\ Most startups struggle to standardize data, so they use a trusted subset. 
\end{tabular}                                                                                       \\ \midrule
Model Learning             & \begin{tabular}[c]{@{}l@{}}A. Arpieg et al., 2018;\\ O. Simeone;\\ D.Fox et al, 2021;\end{tabular}                & \begin{tabular}[c]{@{}l@{}}Startups face challenges in experiment management and troubleshooting deep learning models
\\ They also have the pressure of achieving a lot in a short period, so they might use multiple GPUs to train their model\end{tabular} \\ \midrule
Quality Assurance                & \begin{tabular}[c]{@{}l@{}}L. E. Li et al., 2017;\\ E. d. S. Nascimento et al., 2019;\end{tabular} & \begin{tabular}[c]{@{}l@{}}Startups tend to decompose large ML  models into smaller models to avoid different defects during training \\ As new data is continuously inserted, models should be monitored regularly \end{tabular}                                                                                                                                                                                                                   \\ \midrule
Deployment                & \begin{tabular}[c]{@{}l@{}}L. E. Li et al., 2017;\\ A.Arpteg et al., 2018\\
F. Ishikawa et al., 2019;\end{tabular}                                                                                                                                    & \begin{tabular}[c]{@{}l@{}}ML  models have higher hardware dependency, and different locations might require different models \\ Since most customers of startups do not prefer model versioning, continuous engineering is still elusive. \end{tabular}                                                                                                                                                                                                                                                                                   \\ \bottomrule
\end{tabular}%
  \end{adjustbox}

% }
\caption{Summary of additional SE practices in ML startups}
\label{tab:ml_startup_summary}
\end{table*}

\subsubsection{Testing}

Testing checks if the software does what it is expected to do and ensures the overall quality. Although quality assurance is essential in software development, it is largely absent in most startups \cite{9509046}. 

Pompermaier et al. \cite{pompermaier2017empirical} performed a study on eight startups in Brazil at a tech park. They found that the software tech teams did not use any testing in the first version. So, testing was dependent on the end-users. But on the following version, 75\% of the startups used some testing.

Similarly, Giardino et al. \cite{7360225} found that startups perceive using standard software development practices as a waste of time. So they ignore them to release the products as soon as possible. Therefore, quality assurance is absent in the first versions of the software. Unterkalmsteiner et al. \cite{article2} also concluded that startups generally ignore any quality assurance activities.

Mater et al. \cite{mater2000solving} concluded that startups could outsource their quality assurance test from external experts if the resources are not available in the organization itself. This outsourcing can be an effective alternative for maintaining quality. It can also help in both cost reduction and save time.

Shikta et al.\cite{9509046} based on their study on startups in Bangladesh, proposed a seven-phase framework for quality assurance. The phases are (1) introduce QA lead to achieving quality assurance on requirements and unit testing, (2) lock requirements and regression testing, (3) hire the first dedicated tester, (4) implement test metrics, (5) hire a second dedicated tester to handle new requirements and required bug fixes from customers, (6) hire a third dedicated tester to implement automatic testing, and (7) create a custom testing plan.

Thongsukh et al. \cite{7905012}  suggest that process quality and product quality are closely related to the quality of the software development process. Therefore, using an agile development framework such as scrum methodology will help both startups and big businesses to achieve quality in their product. .

\subsubsection{Deployment}
Software deployment consists of all the activities that involve dispatching products or deliverables to the end-users. It is the final phase of product development. At this phase, the product is ready to be used in a production environment. Startups follow various deployment models. 

Silva et al. \cite{da2005xp} found that some software startups manually deploy the code. While some others use continuous integration tools for deployment, as Taipale et al. \cite{taipale2010huitale} reported. We found that studies about deployment practices are scarce.  

\subsection{RQ2: Which additional software engineering practices do machine learning startups follow?}

Though general software startups share similar software development practices with ML  startups, there are some differences too. Table ~\ref{tab:ml_startup_summary}{} summarizes the additional practices followed by ML  startups. We discussed them in detail in subsequent sections. 

\subsubsection{Requirement Engineering}
At first, ML  startups try to understand problems and define a goal \cite{8870157}. The requirement engineering process followed by general software startups works for ML startups. However, ML  applications have additional requirements that revolve around training data. Alec et al. \cite{banks2019requirements} suggested nine additional considerations for training data requirements. The authors suggested that the training data should be related to high-level requirements, should not contain bias, should be sufficient, self-consistent, reliable, robust, and correct.

However, similar to general software startups, ML  startups face challenges while creating requirements out of business needs \cite{8870157}. The problem is mainly because of the inability of their customers to properly understand what metrics and data are required to address the goal. The challenge is also due to the users' high expectations \cite{10.1145/3461702.3462527}. Customers usually want their products to perform on par with big technology companies like Google. But they cannot meet the requirements due to their data, computing, and expertise limitations.

\subsubsection{Data Management}
Data management is one of the main differences between general software startups and ML  startups. Data is the point where they start their projects for ML  companies. Rob Ashmore et al. \cite{10.1145/3453444} pointed out four activities that entail data management stages. The first activity is data collection which involves either using existing databases or creating a new one. The second activity is preprocessing. In this phase, collected data is adjusted appropriately. The third activity is augmentation. In this phase, the number of data samples is increased using different techniques. And the final step is an analysis of collected and augmented data. Developers can build a data pipeline that deals with structured or unstructured data to build the ML  algorithm. Different kinds of data bugs can appear during the activity mentioned above, and data verification tools can be used by the developers \cite{giray2021software}.
  
Aspen et al. \cite{hopkins2021machine} found that nearly all small companies, including startups, struggled to standardize data entry and collect the correct data, so they tend to rely on a subset of trusted data. However, these companies also found problems with data labeled by the domain expert. And to address that, most of them develop complex routines and reach consensus. Besides that, due to the black-box nature of ML  algorithms, especially neural networks, there is an inherent challenge of data safety, and privacy faced by ML  companies \cite{2018}. 

\subsubsection{Model Learning}
The primary product of a ML  startup is trained ML  models. In the model learning stage, a model or ML  algorithm is trained using training data \cite{8542764}. Rob et al. \cite{10.1145/3453444} suggested four activities in this stage. The first activity is selecting the model that fits from numerous available models. The second activity is training which optimizes the performance of the ML  model. The third activity is hyperparameter selection, which is concerned with choosing the parameters with training activity. The fourth activity suggested by the author is transfer learning which involves reusing ML models across multiple domains.

Startups can face multiple challenges during this phase. Anders et al. \cite{2018} suggested some key challenges for deep learning applications during this phase. The first challenge is experiment management. The author suggested tracking the hardware, platform, source code configuration, training data, and model state during this phase. Troubleshooting a deep learning model is also one of the major challenges, as it is difficult to estimate the results before a system has been fully trained. Moreover, startups have the pressure of achieving a lot of progress in a short period. Dylan et al. \cite{WinNT} suggested training and retraining models quickly. Startups can train on more GPUs or train with lower precision. 

\subsubsection{Quality assurance}
One of the main goals of ML  applications is to make predictions and inferences which make ML startups different from general startups \cite{8411764}. Rob et al. \cite{10.1145/3453444} suggested three activities for this phase. The first activity is requirement encoding, which involves transforming requirements to verifiable test and mathematical properties. The second one is test-based verification which consists of providing test cases to the trained model and checking whether it outputs the expected outcome. The third one is formal verification which is carried out by mathematical techniques to provide sufficient evidence that the model satisfies the requirements. 

While training a ML  model multiple times on the same data, different defects can be observed \cite{hopkins2021machine}. Aspen et al. \cite{hopkins2021machine} pointed out that big companies like Google and Microsoft mitigate the problem by training the model multiple times, but small companies and startups might not afford the resource required to do so. In such cases, developers decompose larger models into smaller models. Besides, the trained model should behave the same way in both real-time mode and batch mode \cite{li2017scaling}. Moreover, new data is continuously inserted into the ML models in a production environment, so developers should continue monitoring model performance \cite{8870157}.

\subsubsection{Deployment}
Rob et al. \cite{10.1145/3453444} suggested three activities in the deployment phase of ML applications. The first is integration, which involves integrating the ML models to wider system architecture. The second is monitoring. It involves monitoring inputs provided to the model, monitoring the environment, monitoring internals of the model, and monitoring outputs of the model. Finally, the deployed model should be updated regularly since distribution of data changes as time passes. However, updating and model versioning is still elusive for many startups \cite{hopkins2021machine} as continuous engineering of imperfect ML  applications might not be favored by customers \cite{8836142}.

Unlike general software, ML applications have a higher dependency on hardware as the ML models' performance depends on GPUs \cite{arpteg2018software}. Besides that, various models need to be deployed in different locations to address the customers \cite{li2017scaling}.

%% file: sec_threats.tex
We discuss the limitations of our study in this section. 

\textbf{Construct Threat}: We only considered papers from January 1, 2000, to November 30, 2021, for this study. This timeframe excludes all the papers published before and after this date, limiting our scope. Besides, we only considered papers published in journals or conferences. We considered a few blogs related to ML  startups as papers on that topic were quite rare. Also, as our priority database was IEEE Xplore, we mostly used papers from that source which could have created a bias. Besides that, we ignored all the gray literature like news articles, videos, podcasts, interviews, etc. But we see the risk associated with this study minimal as most startups do not publish papers explaining their software engineering.  

\textbf{External Threat}: We only used 37  papers for our study. Though that number helped derive insight on software engineering practices for general and ML software startups, it is still a small sample. Also, as papers specifically mentioning startups were rare, especially for ML  startups, we used papers about small teams or companies. Being a small company or having a small team is a startup feature, but it does not always represent startups. So, our findings might not be generalizable to all the startups.  

\textbf{Internal Threat}: The primary internal validity for our study comes from our usage of a pool of papers which we got from specific use of keywords and searching in reliable databases like IEEE Xplore and ACM Digital Library. We further validated the papers by automatically removing papers that do not contain all the keywords in the abstract and then manually verifying them by reading abstracts of the remaining papers. We also enriched the database of papers from Google Scholar and ml-ops.org. Finally, we also use snowballing and manual search to add relevant papers to our collection. Therefore we find this threat minimal.  

%% file: sec_discussion.tex
Our study is the first to do a comparative study of software engineering practices between general software startups and machine learning startups. Our goal was to find out the software engineering practices that predicted the success of startups. It would be beneficial for new startups to follow one such guideline that is more rigorous and proven. General software startups would develop confidence in using those techniques. But, our study showed that there is no such process that predicted success, especially in terms of revenue and longevity of the company. On the other hand, we found that using software engineering practices improved the overall quality of work and employee satisfaction. In the long run, using proper software engineering practices helped better deliver software. 

Machine learning startups share similar software engineering practices with general startups and follow additional practices to address its peculiar challenges. The primary difference comes from the data required for machine learning applications. This difference results in additional practices during the requirement engineering and data management phase. Similarly, the development phase followed by the general startup is replaced by the model learning phase. Both general and machine learning startups have time pressure to deliver respective products quickly with limited resource availability. Therefore, machine learning startups tend to decompose large models into smaller ones and use multiple GPUs to meet the deadline.

%Most of the literature for software startups  we used was based on surveys done all around the globe. One could question the effectiveness of such surveys. Also, most of the studies were geographically distributed and could be heavily  influenced by local culture. This also challenged the generalization of the result. Also, for startups the budget would be the main concern and depending on the labour market and maturity of the technology in the particular country, there could be huge challenges in hiring software engineering professionals and following proper software engineering practices. 

We also aimed to find the tools used in each phase of the software development process. Unfortunately, our goal was hindered by the lack of study in this area. In some literature, we found that using tools helped enhance the software development process. However, we did not find enough information to suggest an exhaustive list of tools that could be used by startups.

In the future, we plan to conduct case studies on selected general and machine learning startups to understand the difference in their software engineering practices as papers based on startups are limited. Interviewing developers from different startups could also be considered to gain insight into how these startups implement software engineering practices during various stages of SDLC. In addition to that, to mitigate the lack of papers, gray literature like interviews of startups founders, videos, and podcasts, medium blog posts could be good resources to understand current software engineering practices adopted by startups.